\documentclass[twocolumn,prl]{revtex4}

\usepackage{amsmath}
\usepackage{amstext}
\usepackage{latexsym}
\usepackage{psfig}
\usepackage{graphicx}
\usepackage{amsfonts}
\newcommand{\ket}[1]{\left\vert#1\right\rangle}
\newcommand{\bra}[1]{\left\langle#1\right\vert}
\begin{document}


  \title{Conditional Production of Superpositions of Coherent States
 with Inefficient Photon Detection}
  \author{A.P. Lund,$^{1,*}$ H. Jeong,$^{1,}$\footnote{The first two authors 
contributed equally to this work.}
 T.C. Ralph,$^1$ and M.S. Kim$^2$}
  \affiliation{$^1$Center for Quantum Computer Technology, 
Department of Physics, University of Queensland, St Lucia, Qld 4072,
   Australia \\
$^2$School of Mathematics and Physics, Queen's University, Belfast BT7 1NN,
   United Kingdom}
    \date{\today}  
\begin{abstract}
It is shown that a linear superposition of two macroscopically
distinguishable optical coherent states can be generated using a single
photon source and simple all-optical operations. Weak squeezing on
a single photon, beam mixing with an auxiliary coherent state, and
photon detecting with imperfect threshold detectors are enough to generate 
a coherent state superposition 
in a free propagating optical field with a large
coherent amplitude ($\alpha>2$) and high fidelity ($F>0.99$). 
In contrast to all previous schemes to generate such a state,
our scheme does not need photon number resolving measurements nor
Kerr-type nonlinear interactions. Furthermore, it is robust to detection
inefficiency and exhibits some resilience to photon production inefficiency.
\end{abstract}

  \maketitle

Since Schr{\" o}dinger suggested his famous cat paradox \cite{Schr},
there has been great interest in generating and observing a quantum
superposition of a macroscopic system. The component states 
composing such a superposition state should be macroscopically distinguishable,
 {\it i.e.}, they should give macroscopically distinct  measurement outcomes
\cite{Reid}. Two coherent states can be discriminated by a homodyne measurement,
which can be considered  a macroscopic measurement, when they are well separated
in the phase space. Therefore, a superposition of two optical coherent states
with sufficiently large amplitudes and with a $\pi$
phase difference between these amplitudes is considered a realization of 
such a macroscopic superposition. 

Recently, the coherent state superposition (CSS) in a free propagating optical
 field  has been studied  for application to
quantum information processing including quantum teleportation \cite{Enk,JKL01},
 quantum computation \cite{JK,Ralph,Ralph2},
entanglement purification \cite{JKpuri} and error correction \cite{Glancy}.
In particular, it was found that quantum computation can
be realized using only linear optics and photon counting, 
given pre-arranged CSS's \cite{Ralph,Ralph2}.
 In this framework, initial CSS's of amplitude  $\alpha \geq  2$ are 
required for efficient quantum computation with simple optical
networks \cite{Ralph2}.

Unfortunately, it is extremely demanding to generate a 
free propagating CSS using current technology.
It is well known that the CSS can be generated from a
coherent state by a nonlinear interaction in a Kerr medium
\cite{Yurke}. However, Kerr nonlinearity of currently available
nonlinear media is extremely small and attenuation is not negligible
compared with the required level to generate a CSS \cite{Boyd}.

Some alternative methods have
been studied to generate a superposition of macroscopically
distinguishable states based upon conditional measurements
\cite{Song,Dakna}. A crucial drawback of these schemes
 is that a highly efficient detector which can discriminate  
photon numbers is necessary. 
Cavity quantum electrodynamics has been studied to
enhance nonlinearity \cite{Tu}.
Some success has been reported in creating such superposition states within
high Q cavities in the microwave \cite{MB} and optical \cite{Mon} domains.
 However, all the suggested schemes for quantum
information processing with coherent states
\cite{Enk,JKL01,JK,Ralph,Ralph2,JKpuri,Glancy} 
require a {\it free propagating} CSS.

In this Letter, we show that a free propagating optical CSS
can be generated with a single photon source and simple optical operations.
A CSS with a small coherent amplitude ($\alpha\leq1.2$) and high fidelity
($F>0.99$) can be deterministically generated by squeezing a single
photon. A large CSS ($\alpha > 2$) with high fidelity ($F>0.99$)
 can be obtained in a non-deterministic way from small CSS's.
Weak squeezing, beam mixing with an auxiliary coherent field and photon
detecting  with threshold detectors are enough to generate such large CSS's
 given a single photon source. 
 Remarkably, neither discrimination
of photon numbers nor $\chi^{(3)}$ nonlinear interactions are required in our
scheme. Furthermore, our scheme is
robust to detection inefficiency
and somewhat resilient to photon production inefficiency.
In a more general sense, our examples reveal some previously unrealized 
relations between the quantum states of harmonic oscillators: we learn that the first 
excited energy eigenstates can be converted to superpositions of large coherent
states by linear operations and projections.

A CSS can be defined as
$|{\rm CSS}_\varphi(\alpha)\rangle=
N_\varphi(\alpha)(|\alpha\rangle+e^{i\varphi}|-\alpha\rangle)$,
where $N_\varphi(\alpha)$ is a normalization factor,
$|\alpha\rangle$ is a coherent state of amplitude $\alpha$,
and $\varphi$ is a real local phase factor. 
The amplitude $\alpha$ is assumed to be real for simplicity without loss of generality.  
In this paper we refer to the magnitude of $\alpha$ as the size of the CSS.  
Note that CSS's  such as 
$|{\rm CSS}_\pm(\alpha)\rangle=N_{\pm}(\alpha)(|\alpha\rangle\pm|-\alpha\rangle)$
are called even and odd CSS's respectively because the even (odd) CSS, 
$|{\rm CSS}_+(\alpha)\rangle$ ($|{\rm CSS}_-(\alpha)\rangle$),
always contains an even (odd) number of photons.

\begin{figure}
\centerline{\scalebox{0.5}{\includegraphics{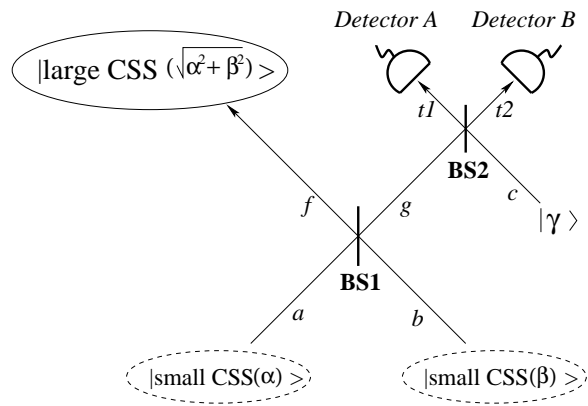}}}
\caption{
A schematic of the non-deterministic CSS-amplification process.
See text for details.}
\label{fig-1}
\end{figure}

{\it An arbitrarily large CSS can be produced
out of arbitrarily small CSS's using the simple experimental set-up 
depicted in Fig.~\ref{fig-1}.}
Let us first illustrate this procedure with a simple example.
 Suppose that one has a collection of identical small odd CSS's with known 
amplitude $\alpha_i$.
Two of the small CSS's are selected and incident onto a 
50:50 beam splitter BS1 as
\begin{widetext}
\begin{equation}
|CSS_-(\alpha_i)\rangle_a|CSS_-(\alpha_i)\rangle_b\stackrel{\rm BS1}{\longrightarrow}
|0\rangle_f\Big(|\sqrt{2}\alpha_i\rangle_g+|-\sqrt{2}\alpha_i\rangle_g\Big)
-\Big(|\sqrt{2}\alpha_i\rangle_f+|-\sqrt{2}\alpha_i\rangle_f\Big)|0\rangle_g
\label{re-re-1}
\end{equation}
\end{widetext}
where the normalization factor is omitted on the right hand side.
One can then say that if one could condition on detecting $|0\rangle_g$,
a larger CSS with amplitude $\sqrt{2}\alpha_i$ would be obtained at mode $f$. 
An additional step is therefore needed to unambiguously discriminate between the vacuum and
coherent states $|\pm\sqrt{2}\alpha_i\rangle_g$ with inefficient detectors. 
Another 50:50 beam splitter, BS2, mixes the field at mode $g$ and an auxiliary coherent state  
$|\sqrt{2}\alpha_i\rangle_c$ as
\begin{widetext}
\begin{equation}
|{\rm BS1}\rangle_{f,g}|\sqrt{2}\alpha_i\rangle_c
\stackrel{\rm BS2}{\longrightarrow} 
|0\rangle_f\Big(|2\alpha_i\rangle_{t1}|0\rangle_{t2}+|0\rangle_{t1}
|2\alpha_i\rangle_{t2}\Big)-\Big(|\sqrt{2}\alpha_i\rangle_f
+|-\sqrt{2}\alpha_i\rangle_f\Big)|\alpha_i\rangle_{t1}|-\alpha_i\rangle_{t2}
\label{re-re-2}
\end{equation}
\end{widetext}
where $|{\rm BS1}\rangle_{f,g}$ represents the right hand side of Eq.~(\ref{re-re-1}) and
the normalization factor is omitted.
Finally, photodetectors $A$ and $B$ are set to detect photons 
at modes $t1$ and $t2$.
The remaining state at mode $f$ is selected
only when both the detectors detect any photon(s) at the same time.
In this case, it is obvious that the right hand side of Eq.~(\ref{re-re-2})
is reduced to a larger CSS's. 
If either of the detectors fails to click, the resulting state is discarded.
This process can be recursively applied until
 a sufficiently large CSS is obtained. Suppose that
an even CSS with amplitude $\alpha\geq2$ is required
while the initial amplitude of small odd CSS's 
is $\alpha_i=1/\sqrt{2}$.
After a sufficient number of CSS's of the amplitude $\sqrt{2}\alpha_i$
are obtained,
the second step will be taken with the same experimental set-up 
and another auxiliary coherent state $|2\alpha_i\rangle$.
In this second stage, larger even CSS's of amplitude $2\alpha_i$ 
will be gained from pairs of even CSS's of $\sqrt{2}\alpha_i$.
 Eventually, the amplitude will reach the required value  by
 four recursive applications of the process, {\it i.e.},
    $\alpha=4\alpha_i\approx 2.83$.

\begin{figure}
\centerline{\scalebox{0.5}{\includegraphics{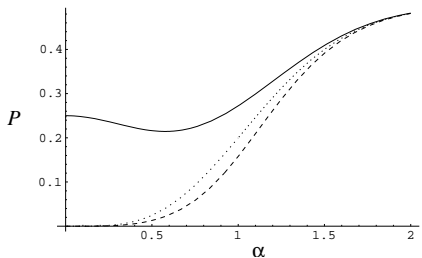}}}
\caption{The success probabilities of the CSS-amplifying process in
Fig.~\ref{fig-1} for the input fields of two identical odd CSS's (solid line),
two identical even CSS's (dashed line), and even and odd CSS's (dotted line).}
\label{fig:prob}
\end{figure}

The process described above can be generalized for arbitrarily small CSS's with 
known amplitudes as shown in Fig.~1.
Suppose two small CSS's, 
$|{\rm CSS}_\varphi(\alpha)\rangle$ and $|{\rm CSS}_\phi(\beta)\rangle$,
with amplitudes $\alpha$ and $\beta$. 
The reflectivity $r$ and transmitivity $t$ of BS1
 are set to $r=\beta/\sqrt{\alpha^2+\beta^2}$ and 
$t=\alpha/\sqrt{\alpha^2+\beta^2}$, where 
the action of the beam splitter is represented by
${\hat B}_{a,b}(r,t)|\alpha\rangle_a|\beta\rangle_b=
|t\alpha+r\beta\rangle_f|-r\alpha+t\beta\rangle_g$.
The other beam splitter BS2 is a 50:50 beam splitter ($r=t=1/\sqrt{2}$)
regardless of the conditions and
the amplitude $\gamma$ of the auxiliary coherent field is 
determined as $\gamma=2\alpha\beta/\sqrt{\alpha^2+\beta^2}$.
The resulting state for mode $f$ then becomes 
$|{\rm CSS}_{\varphi+\phi}({\cal A})\rangle\propto|{\cal A}\rangle+e^{i(\varphi+\phi)}|-{\cal A}\rangle$,
whose coherent amplitude ${\cal A}=\sqrt{\alpha^2+\beta^2}$ is larger than both $\alpha$ and $\beta$.
The relative phase of the resulting CSS
is the sum of the relative phases of the input CSS's.
The success probability $P_{\varphi,\phi}(\alpha,\beta)$ for a single iteration 
is 
\begin{equation}
P_{\varphi,\phi}(\alpha,\beta)
=\frac{(1-e^{-\frac{2\alpha^2\beta^2}{\alpha^2+\beta^2}})^2
[1+\cos(\varphi+\phi)e^{-2(\alpha^2+\beta^2)}]}
{2(1+\cos\varphi e^{-2\alpha^2})(1+\cos\phi e^{-2\beta^2})},\nonumber
\end{equation}
which is plotted for a number of different combinations 
in Fig.~\ref{fig:prob}.
The success probability approaches 1/2 as the amplitudes of initial CSS's
becomes large. Note that the probabilities depend on the type of CSS's
(odd or even) used. The effect of detector inefficiency is just to decrease
this success probability.

\begin{figure}
\centerline{\scalebox{0.5}{\includegraphics{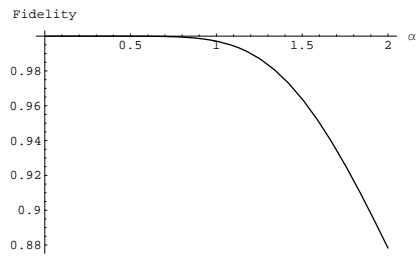}}}
\caption{The fidelity between an odd CSS and squeezed single photon.
The odd CSS is extremely well approximated by the
squeezed single photon for a small coherent amplitude, $\alpha\leq1.2$. }
\label{fid-nopa}
\end{figure}

{\it We now show that a small odd CSS with $\alpha \leq 1.2$ is surprisingly
well approximated by a squeezed single photon.}  
The single mode squeezing operator is
$\hat{S}(r) = e^{-\frac{r}{2}(\hat{a}^2 - \hat{a}^{\dagger 2})}$,
where $r$ is the squeezing parameter and $\hat{a}$ is the annihilation
 operator.
This operator reduces quantum noise of a vacuum state in the phase 
quadrature by a factor of $e^{-2r}$.  
When the squeezing operator is applied to a single photon the resultant state
can be expanded in terms of photon number states as
\begin{equation}
\hat{S}(r) \ket{1} = \sum_{n=0}^{\infty} \frac{(\tanh r)^n}{(\cosh r)^\frac{3}{2}} 
\frac{\sqrt{(2n+1)!}}{2^n n!} \ket{2n + 1}.
\label{e-fs}
\end{equation}
The state contains only odd photon numbers and has coefficients decaying 
exponentially as $n$ increases in a similar fashion to an odd CSS.  The 
fidelity of this state with an odd CSS is
\begin{equation}
F(r,\alpha) = |\langle CSS_-(\alpha)|S(r)|1\rangle|^2
=\frac{2\alpha^2\exp[\alpha^2(\tanh r-1)]}{(\cosh r)^3(1-\exp[-2\alpha^2])}.
\nonumber
\end{equation}
Fig.~\ref{fid-nopa} shows the maximized fidelity on the y-axis plotted
against a range of possible values for $\alpha$ for the desired odd CSS.
Some example values are:
 $F=0.99999$ for amplitude $\alpha=1/2$,
  $F=0.9998$ for $\alpha=1/\sqrt{2}$, 
and  $F=0.997$ for $\alpha=1$, where the maximizing squeezing parameters are 
$r=0.083$, $r=0.164$, and $r=0.313$ respectively. 
 Firstly note that for $\alpha$
very close to zero the fidelity approaches unity. When $\alpha \rightarrow 0$,
$r \rightarrow 0$ and hence the squeezing operator $\hat S(r)$ approaches
the identity transformation. An odd CSS with $\alpha$ very
close to zero has a significant contribution from a single photon 
and very little
from higher odd photon numbers.  This is the reason for the high fidelity as
$\alpha$ tends to zero.  The fidelity remains high for $\alpha$ near zero as
one can match the three photon contribution to the CSS by the
squeezing operator whilst still being able to neglect higher order 
photon number
terms.  Eventually as $\alpha$ increases, higher photon numbers cannot be
matched and so as $\alpha$ tends to infinity, the fidelity tends to zero.

As the fidelity between
a squeezed single photon and an ideal small CSS is extremely high,
it can be conjectured that 
a large CSS distilled from squeezed single photons by our scheme will 
also be very close to an ideal large CSS.
 In what follows, we will show that this conjecture is true for $\alpha\leq2.5$.

In order to calculate multiple iterations we need to use numerical techniques.
 We are using coherent states of
 some bounded coherent amplitude and superpositions there of. Provided the
 coherent amplitudes are not too large, the most significant contributions
 to these states are Fock states of low number.  For computations here 
the lowest thirty Fock states were used.  This provides a very good
 approximation for coherent states with $\alpha \leq 2.5$.  
All 29 possible ``click'' events are included 
for all detectors.   

\begin{figure}
\centerline{\scalebox{0.5}{\includegraphics{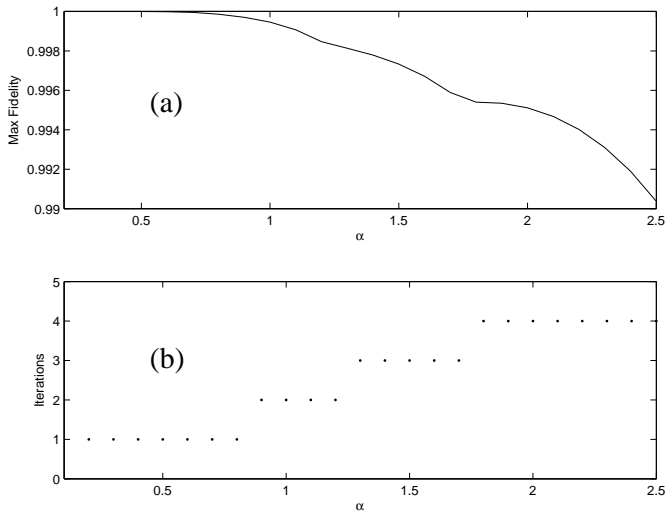}}}
\caption{(a) The maximum fidelity obtained in our scheme vs the coherent
 amplitude. (b) The number of iterations which gives the maximum fidelity vs
 the coherent amplitude.}
\label{something}
\end{figure}

If one wished to create a CSS with a particular $\alpha$ with $n$ 
CSS-amplification steps, then initial CSS's with 
$\alpha_i = \alpha/\sqrt{2}^n$ are required.
 As the number of steps increases the required $\alpha_i$ decreases.  
 When generating a large CSS 
out of the squeezed single photon states 
  the fidelity maximizes for
 a particular number of iterations.  Fig.~\ref{something} shows the maximum
 possible fidelity using this process in (a) and the number of steps
 in (b) against the desired $\alpha$ in the CSS's.
For example, four iterations starting from the initial amplitude $\alpha_i=1/2$ is required to
gain the maximum fidelity $F=0.995$ for $\alpha=2$.
It is evident from Fig.~4 that high fidelity, $F>0.99$, can be obtained up to $\alpha= 2.5$.
The error rate for discrimination between coherent states with $\alpha=\pm 2.5$
via a classical measurement (homodyne detection) is only $3\times10^{-7}$.

Current technology does not produce pure single photon states;
the single photon is always in a mixture with the vacuum as 
$p\ket{0}\bra{0}+ (1-p)\ket{1}\bra{1}$, where $p$ is the inefficiency of the photon production.
Hence the squeezed single photon state will also be a mixture with a squeezed vacuum.
However, an interesting aspect of our scheme is that
it may be somewhat resilient to the photon production inefficiency because
 its first iteration purifies the mixed CSS's while amplifying them.
The initial input states for the CSS amplification process 
from the imperfect single photon source are
\begin{widetext}
\begin{eqnarray}
&&\rho_{a,b,c}
=\Big[(1-p)^2\ket{S_1}\bra{S_1}\otimes\ket{S_1} \bra{S_1}+p^2
\ket{S_0} \bra{S_0}\otimes\ket{S_0}\bra{S_0}\nonumber\\
&&~~~~~~~~~~~~  + p(1-p) \Big(\ket{S_0} \bra{S_0}\otimes \ket{S_1} \bra{S_1}
  +\ket{S_1} \bra{S_1}\otimes \ket{S_0} \bra{S_0}\Big)\Big]_{a,b}\otimes\big(|\gamma\rangle\langle\gamma|\big)_c,
\end{eqnarray} 
\end{widetext}
where $|S_0\rangle =\hat{S}(r) \ket{0}$ and $|S_1\rangle =\hat{S}(r) \ket{1}$.
Here, the terms with $p^2$ and $p(1-p)$ are undesired error terms where either  (or both) of
the single photons  is missing.
Note that the initial amplitude is required to be small to produce a large CSS with high fidelity.
Provided such a small amplitude,
input states incident onto the beam splitters in our experimental setup
contain approximately only two (or slightly more than two) photons.
In such cases the probability of simultaneous clicks at detectors $A$ and $B$
in Fig.~1 will significantly decrease when any of the single photons is missing.
In other words, the undesired cases will rarely be selected 
for the next iteration of the amplification process.
We have obtained numerical results 
for the initial amplitude $\alpha_i=1/2$ as follows by the methods that we have already
explained. If $p=0.4$, the fidelity of the initial CSS, which is a mixture with a squeezed vacuum,
is $F=0.60$ but it will become $F=0.89$ by the first iteration.
Thus a larger CSS of significantly high fidelity is produced.
If $p=0.25$ $(p=0.05)$, the fidelity of the initial CSS is $F=0.750$ $(F=0.950)$ but
 becomes $F=0.941$ $(F=0.990)$ by the first iteration.
 Such purification effects for multiple iterations are
  beyond the scope of this Letter but deserve further investigations.

In the CSS amplification process, 
the zero amplitude coherent states that occur in the detection modes
in Eq.~(\ref{re-re-2})
 may be slightly different from zero
because of imperfect mode matching 
 at beam splitters.
This will lead to a small probability of accepting the wrong state.
Good mode matching is a requirement in any linear optical network where one wishes
to measure manifestly quantum mechanical effects.
Highly efficient mode matching of a single photon
from parametric down conversion and a weak coherent state from an attenuated laser beam 
at a beam splitter has been experimentally demonstrated
using optical fibers \cite{Pitt}. Such techniques could be employed for
the implementation of our scheme.

The dark count rate of photodetectors will affect the fidelity of the CSS's.
Currently, highly efficient detectors have relatively high dark count
rates while less efficient detectors have very low dark count rates
\cite{Takeuchi}. We emphasize again that our scheme does not require
highly efficient detectors because the inefficiency of the detectors
does not affect the quality of CSS's even though it decreases the success
probability. Silicon avalanche photodiodes operating at the visible wavelength have
relatively high efficiency and a small dark count rate, which is
preferred in our proposal.

The single photons required for our scheme could be generated conditionally
from a down-converter \cite{Lvovsky}. This is a $\chi^{(2)}$ process
(like squeezing) and does not require photon number resolving detection.
Once free propagaing 
CSS's are generated, they can be detected by homodyne
measurements, which can be highly efficient in quantum optics
experiments. 

Our scheme non-deterministically generates large CSS's. 
However, a non-deterministic CSS source is useful enough for quantum
information processing \cite{Ralph,Ralph2}. Efficient gate operations for 
coherent-state quantum computation \cite{Ralph2} are based on
teleportation via an entangled coherent state \cite{Enk,JKL01}. Entangled
coherent states can be simply generated from 
CSS's using a beam splitter and can be used as off-line resources for quantum computation.
We note that such entanglement of macroscopically distinguishable states
is perhaps more closely aligned with Schr\"odinger's original concept \cite{Schr}.

In conclusion, we have proposed a simple all-optical scheme to generate
a linear superposition of macroscopically distinguishable
coherent states in a propagating optical field.
We have found a previously unrealized connection between squeezed number states
and superpositions of coherent states as well as the interesting
additive properties of the latter. 
In stark contrast to all previous schemes our scheme 
requires neither $\chi^{(3)}$ nonlinearity nor photon number resolving detection
to generate a macroscopic superposition state.


We would like to thank A. Gilchrist for useful comments. This work was
supported by the Australian Research Council and the University of
Queensland Excellence Foundation. M.S.K. acknowledges the UK Engineering
and Physical Science Research Councils and the Korea Research Foundation
(2003-070-C00024) for financial support.

\end{document}